\documentclass[a4paper]{jpconf}
\usepackage{graphicx}
\usepackage{psfrag}
\usepackage{color}

\newcommand{\affuni}[2]{Dipartimento di Fisica dell'Universit\`a #1, #2, Italy.}
\newcommand{\affinfn}[2]{INFN Sezione di #1, #2, Italy.}

\begin{document}

\title{\boldmath $\gamma\gamma$ physics with the KLOE experiment}

\author{The KLOE-2 Collaboration}
\author{
F.~Archilli,$^{n,o}$
D.~Babusci,$^f$
D.~Badoni,$^{n,o}$
I.~Balwierz,$^e$
G.~Bencivenni,$^f$
C.~Bini,$^{l,m}$
C.~Bloise,$^f$
V.~Bocci,$^m$
F.~Bossi,$^f$
P.~Branchini,$^q$
A.~Budano,$^{p,q}$
S.~A.~Bulychjev,$^g$
P.~Campana,$^f$
G.~Capon,$^f$
F.~Ceradini,$^{p,q}$
P.~Ciambrone,$^f$
E.~Czerwi\'nski,$^f$
E.~Dan\'e,$^f$
E.~De~Lucia,$^f$
G.~De~Robertis,$^b$
A.~De~Santis,$^{l,m}$
G.~De~Zorzi,$^{l,m}$
A.~Di~Domenico,$^{l,m}$
C.~Di~Donato,$^{h,i}$
D.~Domenici,$^f$
O.~Erriquez,$^{a,b}$
G.~Fanizzi,$^{a,b}$
G.~Felici,$^f$
S.~Fiore,$^{l,m}$
P.~Franzini,$^{l,m}$
P.~Gauzzi,$^{l,m}$
S.~Giovannella,$^f$
F.~Gonnella,$^{n,o}$
E.~Graziani,$^q$
F.~Happacher,$^f$
B.~H\"oistad,$^s$
E.~Iarocci,$^{j,f}$
M.~Jacewicz,$^s$
T.~Johansson,$^s$
V.~Kulikov,$^g$
A.~Kupsc,$^s$
J.~Lee-Franzini,$^{f,r}$
F.~Loddo,$^b$
M.~Martemianov,$^g$
M.~Martini,$^{f,k}$
M.~Matsyuk,$^g$
R.~Messi,$^{n,o}$
S.~Miscetti,$^f$
G.~Morello,$^f$
D.~Moricciani,$^o$
P.~Moskal,$^e$
F.~Nguyen,$^{p,q}$
A.~Passeri,$^q$
V.~Patera,$^{j,f}$
I.~Prado~Longhi,$^{p,q}$
A.~Ranieri,$^b$
P.~Santangelo,$^f$
I.~Sarra,$^f$
M.~Schioppa,$^{c,d}$
B.~Sciascia,$^f$
A.~Sciubba,$^{j,f}$
M.~Silarski,$^e$
C.~Taccini,$^{p,q}$
L.~Tortora,$^q$
G.~Venanzoni,$^f$
R.~Versaci,$^{f,u}$
W.~Wi\'slicki,$^t$
M.~Wolke,$^s$
J.~Zdebik.$^e$}

\begin{center}
$^a${\affuni{di Bari}{Bari}}\\
$^b${\affinfn{Bari}{Bari}}\\
$^c${\affuni{della Calabria}{Cosenza}}\\
$^d${INFN Gruppo collegato di Cosenza, Cosenza, Italy.}\\
$^e${Institute of Physics, Jagiellonian University, Cracow, Poland.}\\
$^f${Laboratori Nazionali di Frascati dell'INFN, Frascati, Italy.}\\
$^g${Institute for Theoretical and Experimental Physics (ITEP), Moscow, Russia.}\\
$^h${\affuni{''Federico II''}{Napoli}}\\
$^i${\affinfn{Napoli}{Napoli}}\\
$^j${Dipartimento di Scienze di Base ed Applicate per l'Ingegneria dell'Universit\`a ``Sapienza'', Roma, Italy.}\\
$^k${Dipartimento di Scienze e Tecnologie applicate, Universit\`a "Guglielmo Marconi", Roma, Italy.}\\
$^l${\affuni{''Sapienza''}{Roma}}\\
$^m${\affinfn{Roma}{Roma}}\\
$^n${\affuni{``Tor Vergata''}{Roma}}\\
$^o${\affinfn{Roma Tor Vergata}{Roma}}\\
$^p${\affuni{``Roma Tre''}{Roma}}\\
$^q${\affinfn{Roma Tre}{Roma}}\\
$^r${Physics Department, State University of New York at Stony Brook, USA.}\\
$^s${Department of Nuclear and Particle Physics, Uppsala Univeristy,Uppsala, Sweden.}\\
$^t${A. Soltan Institute for Nuclear Studies, Warsaw, Poland.}\\
$^u${Present Address: CERN, CH-1211 Geneva 23, Switzerland.}
\end{center}


\begin{abstract}
The processes $e^+e^-\to e^+e^-X$, with $X$ being either
the $\eta$ meson or $\pi^0\pi^0$, are studied
at DA$\Phi$NE, with $e^+e^-$ beams colliding at $\sqrt{s}\simeq1$ GeV,
below the $\phi$ resonance peak. The data sample is from
an integrated luminosity of 240 pb$^{-1}$, collected by the KLOE experiment
without tagging of the outgoing $e^+e^-$. Preliminary results are presented on
the observation of the $\gamma\gamma\to\eta$ process, with both
$\eta\to\pi^+\pi^-\pi^0$ and $\eta\to\pi^0\pi^0\pi^0$ channels,
and the evidence for $\gamma\gamma\to\pi^0\pi^0$ production at low
$\pi^0\pi^0$ invariant mass.
\end{abstract}

\section{Introduction}

The term ``$\gamma\gamma$ physics'' stands for the study of the reaction of order $\alpha^4$
$e^{+} e^{-} \to e^{+} e^{-} \gamma^{*} \gamma^{*} \to e^{+} e^{-} X$, where $X$ is some arbitrary final 
state allowed by conservation laws.
In particular, hadronic states with
$J^{PC} = 0^{\pm+};2^{\pm+}$ are directly produced through the $\gamma^*\gamma^*\to X$ subprocess.
For quasi-real photons the number of produced events can be estimated from the expression
\begin{equation}
N = L_{ee} \int dW_{\gamma\gamma}\frac{dL}{dW_{\gamma\gamma}} \sigma(\gamma \gamma \to X)
\end{equation}
where $L_{ee}$ is the $e^+e^-$ luminosity, $W_{\gamma\gamma}$ the photon-photon center of mass energy ($W_{\gamma\gamma}=M_X$), $dL/dW_{\gamma\gamma}$ the photon-photon flux and $\sigma$ the cross section into a given final state. By
neglecting single powers of $\ln(E/m_e)$, when the scattered leptons are undetected in the
final state one has
\begin{equation}
\frac{dL}{dW_{\gamma\gamma}}=
\frac{1}{W_{\gamma\gamma}}\left(\frac{\alpha}{2\pi}\mbox{ln}\frac{E}{m_e}\right)^2 f(z) \qquad z=\frac{W_{\gamma\gamma}}{2E}
\end{equation}
with $f(z)$ (Low function)~\cite{BKT,Low} given by
\begin{equation}
f(z)= (2+z^2)^2\mbox{ ln}\frac{1}{z}-(1-z^2)(3+z^2) 
\end{equation}
The $\gamma\gamma$ processes studied with KLOE are $e^{+} e^{-} \to e^{+} e^{-} \eta$
and $e^{+} e^{-} \to e^{+} e^{-} \pi^0\pi^0$. Past experiments measuring the
$\gamma\gamma$ cross section for light mesons production, took data at a center of mass energy
$7<\sqrt{s}<35$ GeV. A low energy $e^+e^-$
factory, such as DA$\Phi$NE, compensates the small cross section value
with the high luminosity.

DA$\Phi$NE is an $e^+e^-$ collider designed to operate at the center of mass
energy $\sqrt{s}\simeq1.02$ GeV, namely the $\phi$ meson mass.
The sample used for the present analyses consists of data taken at $\sqrt{s}=1$ GeV, which allows
the reduction of the background from $\phi$ decays, with an integrated luminosity of 240 pb$^{-1}$.

The KLOE detector consists of a large (3.3~m length and 2~m radius) volume drift chamber,
surrounded by a lead--scintillating fibers calorimeter. The detector is inserted in a superconducting coil
producing an axial field $B$=0.52~T. Charged particle momenta are reconstructed with resolution
$\sigma_p/p\simeq0.4\%$ ($\sigma_p/p\simeq1\%$) for large (small) angle tracks.
Calorimeter energy clusters are reconstructed with energy and time resolution of $\sigma_E/E=5.7\%/\sqrt{E({\rm GeV})}$
and $\sigma_t=57\mbox{ ps}/\sqrt{E({\rm GeV})}\oplus 100$ ps.

Data are processed with a dedicated filter asking for
at least two prompt photons, clusters not associated to tracks and propagating with $|\vec{r}-{\rm c}t|\sim0$,
with energy $E>15$ MeV and polar angle $20^\circ<\theta<160^\circ$,
the most energetic with $E>50$ MeV, the fraction of energy carried by photons $R=(\sum_\gamma E_\gamma)/E_{cal}>0.3$ and
the total energy in the calorimeter $100<E_{cal}<900$ MeV. The latter requirement rejects low energy background and the high rate processes $e^{+} e^{-} \to e^{+} e^{-}$, $e^{+} e^{-} \to \gamma \gamma$. 

\section{\boldmath Observation of $\gamma\gamma\to\eta$, with $\eta\to\pi^+\pi^-\pi^0$}
\begin{figure}[htbp]
\psfrag{Miss}{\hspace*{-2.5cm}\tiny M$_{miss}^2$ (GeV$^2$)}
\psfrag{e3afp}{\tiny $E_\gamma$ (MeV)}
\centering
\includegraphics[width = 6cm,height=6.2cm]{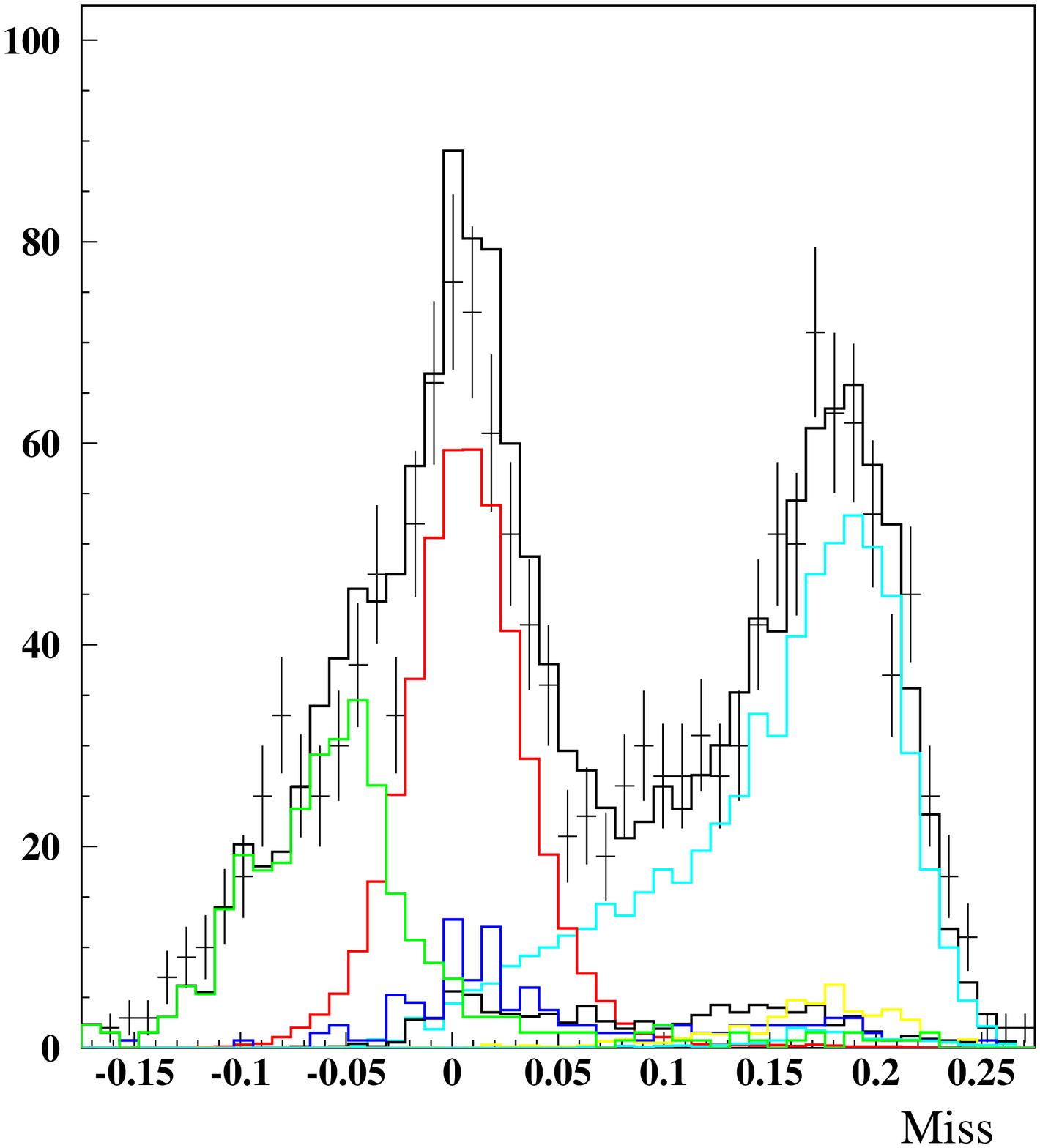}%
\hspace*{2.5cm}
\includegraphics[width = 6cm,height=6.2cm]{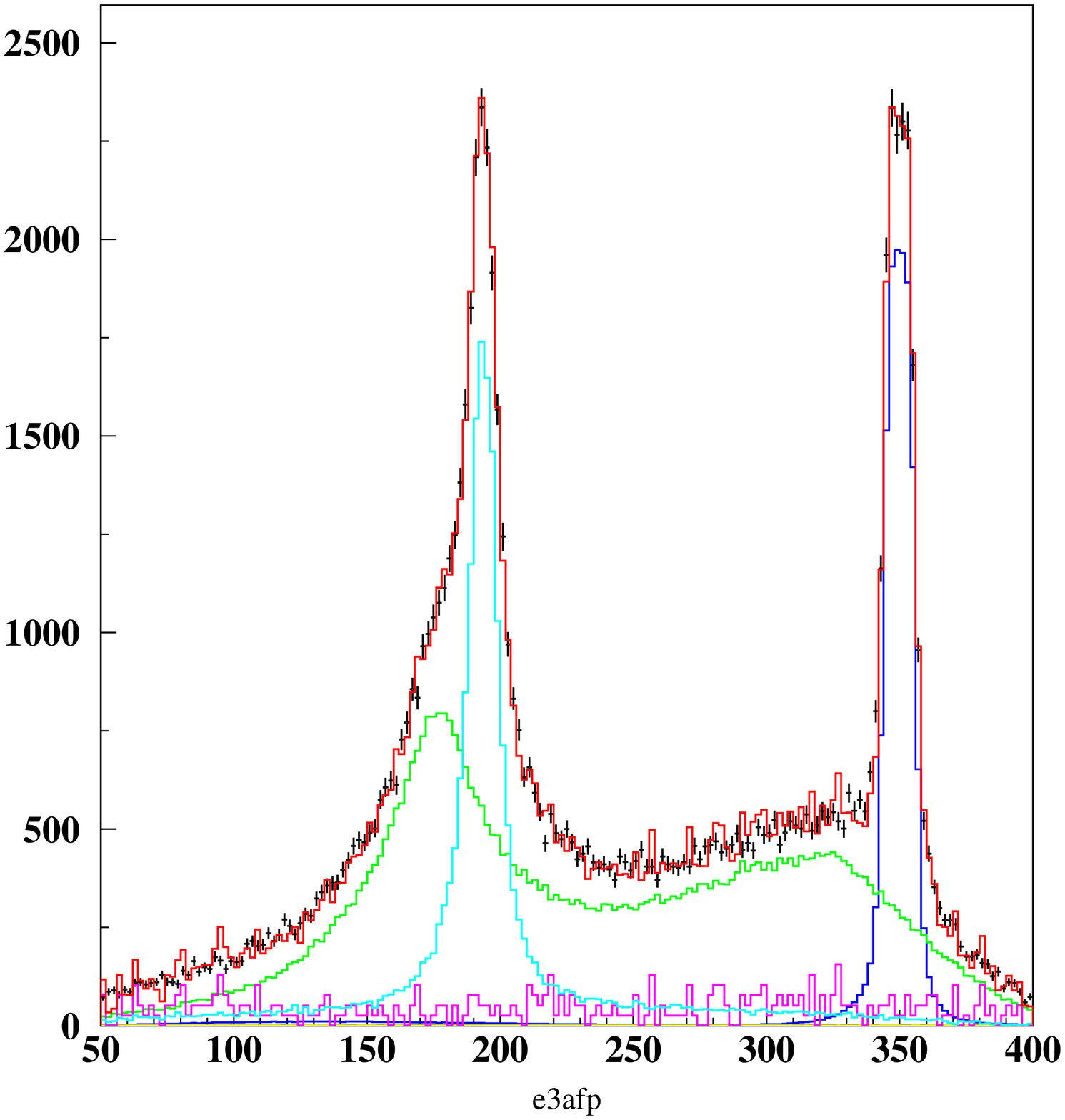}%
\caption{Left: fit of the $M_{miss}^2$ distribution for the $e^+e^-\to e^+e^-\eta$ analysis. Main contributions are: $e^+e^-\gamma$ (green) at negative $M_{miss}^2$ values due to the pion mass assigned to $e^+e^-$ tracks, $\eta\gamma$ (red), signal events (ligt blue). Right: fit of the recoil photon energy spectrum for the $e^+e^-\to\eta\gamma$ analysis. The $\eta\gamma$ peak (blue) and the $\omega\gamma$ peak (light blue) are visible; the broad distribution (green) is due to $\omega\pi^0$ events.}
 \label{f:MPIPIGGpi+pi-pi0}
\end{figure}

The selection of these events asks for two photons compatible with a $\pi^0$
decay and two tracks with opposite curvature coming from the collision point.
The charged pion mass is assigned to the two tracks and a least squares function based on Lagrange multipliers
imposes that $\pi^+\pi^-\pi^0$ come from an $\eta$ decay. Therefore most background events are suppressed, except
for the irreducible process $e^+e^-\to\eta\gamma\to\pi^+\pi^-\pi^0\gamma$, with the monochromatic photon,
$E_\gamma$ = 350 MeV, lost in the beam pipe. For these background events, however, the correlation between the squared missing mass,
$M_{miss}^2$, and the longitudinal momentum, $p_L$, of the $\pi^+\pi^-\pi^0$ system is rather different than for the signal.
Further criteria are applied to suppress processes with photons and $e^+e^-$ in the final state. 
The Monte Carlo, MC, generator used for the signal~\cite{Nguyen:2006sr} allows the full phase space generation
of $e^+e^-\to e^+e^-\gamma^*\gamma^*\to e^+e^-\eta$ events.
Fig.~\ref{f:MPIPIGGpi+pi-pi0} (left) shows the $M_{miss}^2$ distribution for data fitted with the superposition
of MC shapes for signal and background. An independent fit is performed with the $\pi^+\pi^-\pi^0$
longitudinal momentum. Both fits show the same yields for the background processes and more than 600 signal events. 
The cross section of the irreducible process $e^+e^-\to\eta\gamma\to\pi^+\pi^-\pi^0\gamma$
is evaluated using the same sample of data and 
asking for three photons in the final state. A kinematic fit is performed, requiring energy and momentum conservation,
and the improved variables are used to fit in the distribution of the recoil photon energy (Fig.~\ref{f:MPIPIGGpi+pi-pi0}, right).
The preliminary result is $\sigma_{e^+ e^-\to\eta\gamma}(1\mbox{ GeV}) = 0.866\,(9)_{stat}\,(93)_{syst}$ nb,
where the systematic uncertainty is given by residual $e^+e^-\to\omega\pi^0\to\pi^+\pi^-\pi^0\pi^0$ background.
The cross section $\sigma_{\gamma\gamma\to\eta\to\pi^+\pi^-\pi^0}(1\mbox{ GeV})$ is under evaluation.

\section{\boldmath Observation of $\gamma\gamma\to\eta$, with $\eta\to\pi^0\pi^0\pi^0$} 
\begin{figure}[htbp]
\centering
\includegraphics[width = 6.5cm,height=6.2cm]{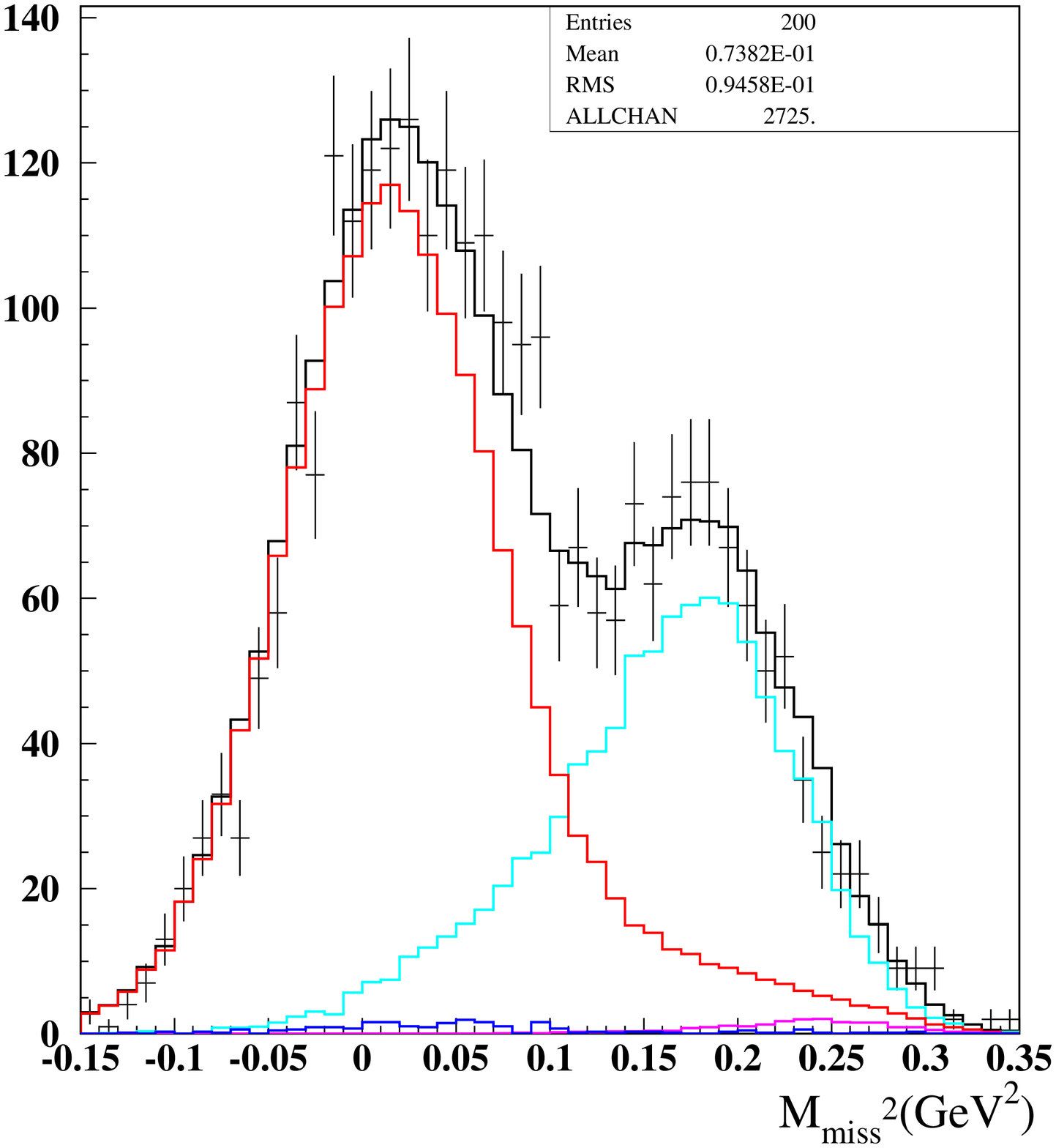}  
\hspace*{2cm}
\includegraphics[width = 6.5cm,height=6.2cm]{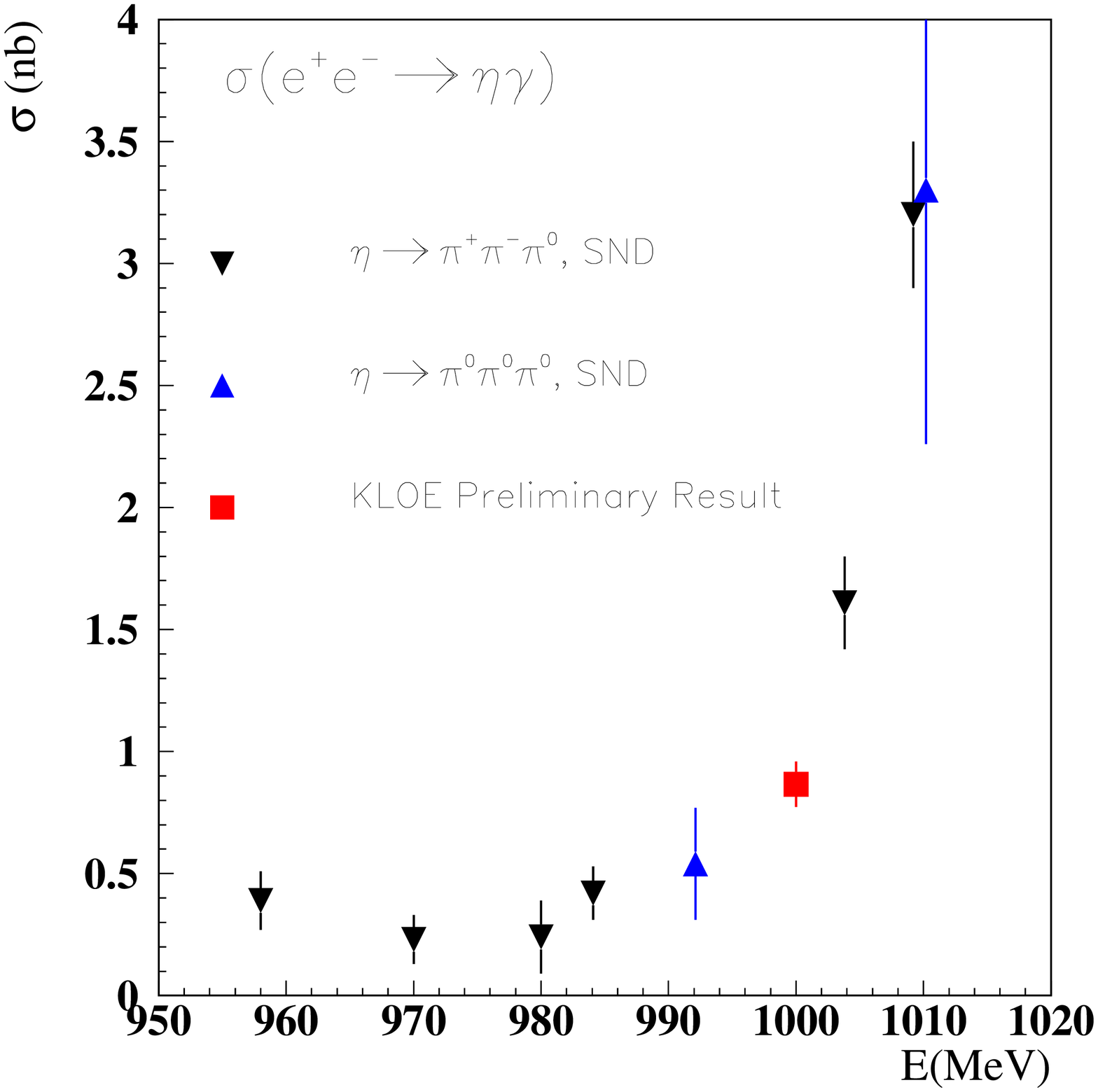}
\caption{$\gamma\gamma\to\eta\to\pi^0\pi^0\pi^0$ analysis.
Left: squared missing mass distributions for data (points whith error bars) and backgrounds (histograms) normalized 
according to the fit results. Colour code: red = $e^+e^-\to\eta\gamma$, light blue = signal, blue = $e^+e^-\to\omega\pi^0$. 
Right: preliminary KLOE value for $\sigma(e^+e^-\to\eta\gamma)$ at $\sqrt{s}=1\mbox{GeV}$ (red point), and SND 
results~\cite{ach} for several values of $\sqrt{s}$.}
\label{fig:eta}
\end{figure}

The main backgrounds for this analysis are annihilation processes with at least four 
prompt photons in the final state, $e^+e^-\to\eta\gamma,\mbox{ }K_SK_L,\mbox{ }\omega\pi^0$,
for which accidental or split calorimeter clusters increase the photon multiplicity.
Events with six and only six prompt photons in the final state and with no tracks in the drift chamber are selected.
The photons are paired choosing the combination which minimizes the $\chi^2$-like variable
\begin{equation}
\chi^2_{6\gamma}=\frac{\left(m_{\pi^0}-m_{ij}\right)^2}{\sigma^2_{ij}}+\frac{\left(m_{\pi^0}-m_{mn}\right)^2}{\sigma^2_{mn}}+
\frac{\left(m_{\pi^0}-m_{kl}\right)^2}{\sigma^2_{kl}},
\label{eq:chipair}
\end{equation}
where $m_{ij}$ is the invariant mass of each pair of photons and $\sigma_{ij}$ the resolution.
Then, a kinematic fit is performed asking for the six photons invariant mass to be equal to
the mass of the $\eta$ meson. A cut is then applied on the energy of the most energetic photon,
to reject $e^+e^-\to\eta\gamma$ events in which the monochromatic photon is detected. 
The squared missing mass, $M_{miss}^2$, distribution is well described by the $\gamma\gamma\to\eta\to\pi^0\pi^0\pi^0$ and
$\eta(\to\pi^0\pi^0\pi^0)\gamma$ MC shapes. Fig.~\ref{fig:eta}, left, shows the fit in $M_{miss}^2$.

From the fit we obtain about 900 $\gamma\gamma\to\eta\to3\pi^0$ events and
a preliminary value for the cross section, measured with the $\eta\to\pi^0\pi^0\pi^0$ channel,
$\sigma_{e^+ e^-\to\eta\gamma}(1\mbox{ GeV}) = 0.875\,(18)_{stat}\,(35)_{syst}$, in agreement
with that measured through the $\eta\to\pi^+\pi^-\pi^0$ channel.
This is shown in Fig.~\ref{fig:eta} (right) among other experimental results~\cite{ach} as a function of $\sqrt{s}$.
We are finalizing the evaluation of the $\sigma_{\gamma\gamma\to\eta\to\pi^0\pi^0\pi^0}(1\mbox{ GeV})$ cross section.

\section{\boldmath Search for $\gamma\gamma\to\pi^0\pi^0$}
The main goal of this analysis is to investigate the low $\pi^0\pi^0$ invariant mass region, just above the production threshold,
where a contribution from the $\sigma(600)$ scalar
meson~\cite{Aitala:2000xu,Muramatsu:2002jp,Ablikim:2004qna} as an intermediate state is expected. 

\begin{figure}
\centering
\includegraphics[width = 6.5cm,height=6.2cm]{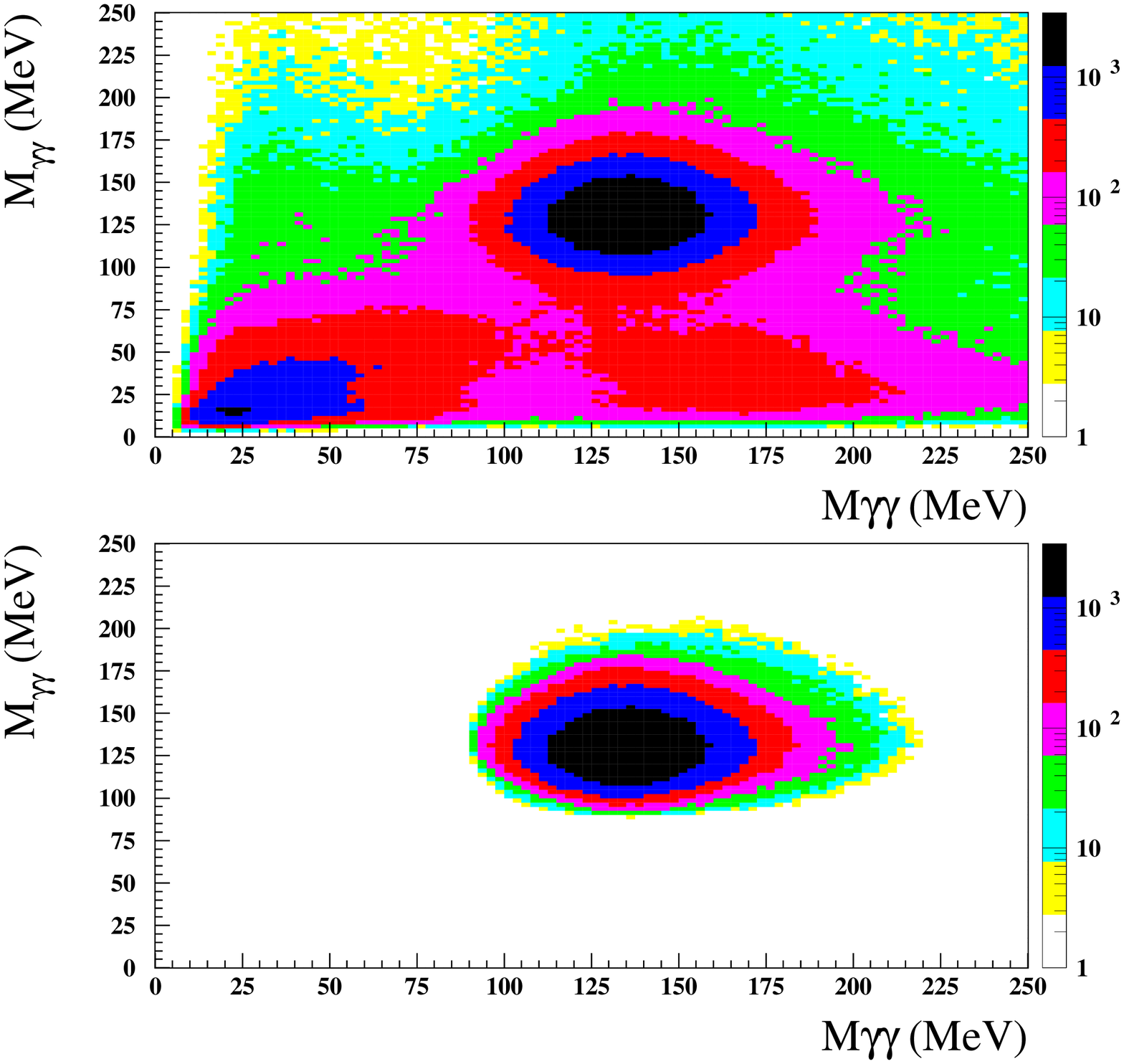}  
\hspace*{2cm}
\includegraphics[width = 6.5cm,height=6.2cm]{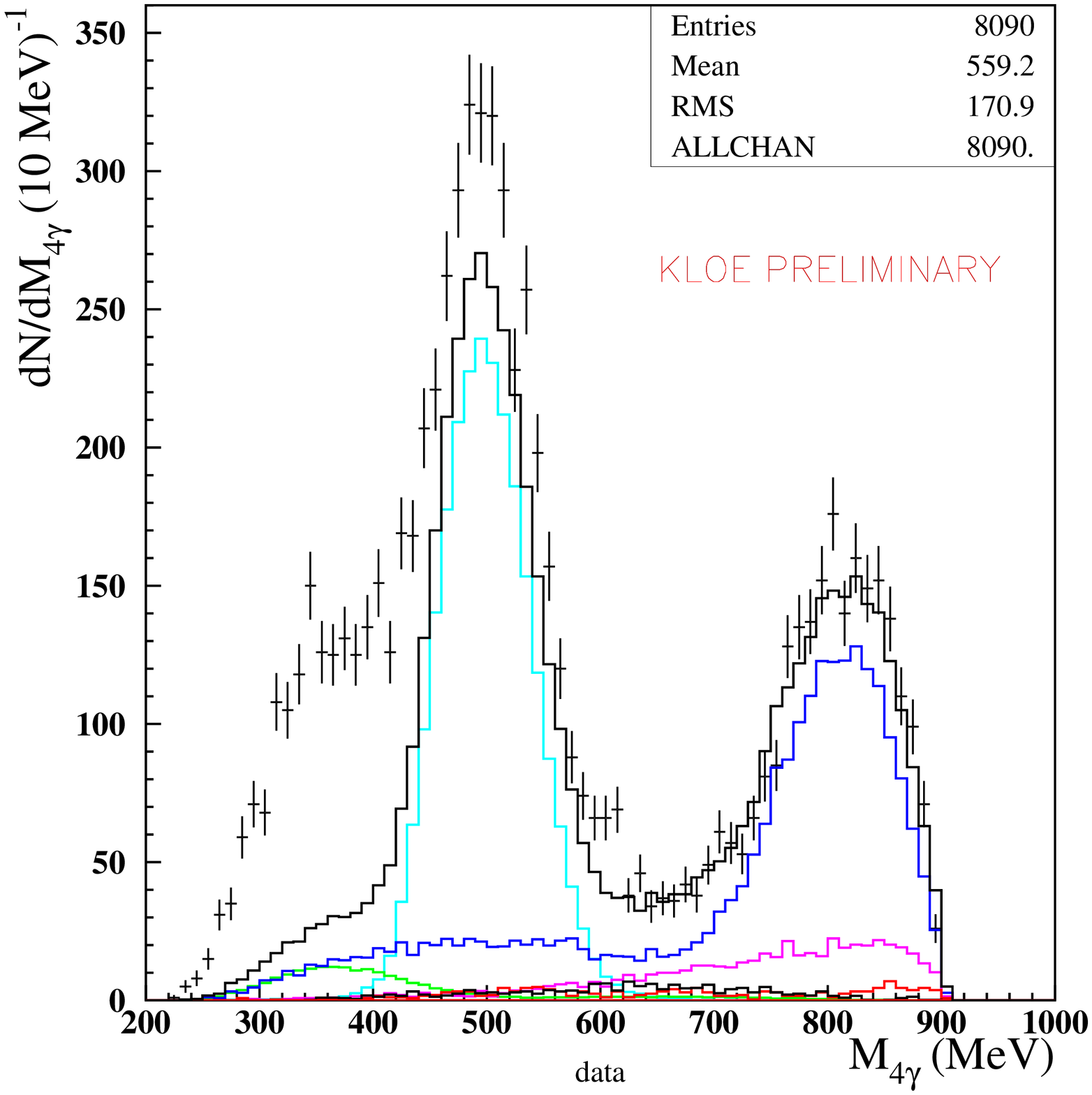}
\caption{$\gamma\gamma\to\pi^0\pi^0$ analysis. Left: scatter plots in two photons pairs invariant masses before (up) and after (down)
 rejecting bad $\chi^2_{4\gamma}$ events; selected events have both photons pairs invariant masses centered around $\pi^0$ mass value. 
Right: 4$\gamma$ invariant mass spectrum for data (points with error bars) and backgrounds. 
Colour code: light blue = $e^+e^-\to K_SK_L$, blue = $e^+e^-\to\omega\pi^0$, violet =
$e^+e^-\to f_0(980)\gamma$, green = $e^+e^-\to\eta\gamma$, red = $e^+e^-\to\gamma\gamma$.}
\label{fig:pi0pi0}
\end{figure}

The main backgrounds are annihilation processes with four or more prompt photons in the final state:
$e^+e^-\to K_SK_L,\eta\gamma,\omega\pi^0,f_0(980)(\to\pi^0\pi^0)\gamma$. In addition, 
due to the possibility of cluster splitting, the $e^+e^-\to \gamma\gamma$ process is also considered as a source of background. 
Selected events have no tracks in the drift chamber and four prompt photons in the final state with polar angle 
$23^{o}<\vartheta<157^{o}$ and energy $E_{\gamma}>15\mbox{ MeV}$.
The photons are paired choosing the combination which minimizes the $\chi^2$-like variable
of eq.(\ref{eq:chipair}) in the case of four photons coming from two neutral pions, $\chi^2_{4\gamma}$.
Events with $\chi^2_{4\gamma}>4$ are rejected: the effect of this selection is shown in
Fig.~\ref{fig:pi0pi0} (left). 
To reject $e^+e^-\to K_SK_L$ events, where a large amount of non-prompt energy is released in the detector
a tighter cut on the energy fraction carried by photons, $R=(\sum_\gamma E_\gamma)/E_{cal}>0.75$, is applied.
The four photon invariant mass spectrum for the selected data sample is shown in Fig.~\ref{fig:pi0pi0} (right) 
together with the normalized background Monte Carlo simulations. 
The excess of events with respect to the expected annihilation processes,
in the low invariant mass region, is an indication of the $\gamma\gamma\to\pi^0\pi^0$ production. 
The determination of the differential cross section ${\rm d}\sigma_{\gamma\gamma\to\pi^0\pi^0}/{\rm dM}_{4\gamma}$
is in progress.

\section{Conclusions}
From an integrated luminosity of 240 pb$^{-1}$ of data
collected at DA$\Phi$NE, operating at $\sqrt{s}\simeq1$ GeV, the
following preliminary results in $\gamma\gamma$ analyses are achieved:
\begin{itemize}
\item unambiguous signature of both $\gamma\gamma\to\eta$ and
  $\gamma\gamma\to\pi^0\pi^0$ events, without any $e^\pm$ tagger;
\item $\gamma\gamma\to\eta$ events are observed through both
  $\eta\to\pi^+\pi^-\pi^0$ and $\eta\to\pi^0\pi^0\pi^0$ channels, with independent
  systematics;
\item from the same data sample, an exploratory research shows a structure at
  small values of the $M_{4\gamma}$ spectrum, where the process $e^+e^-
  \to e^+e^-\pi^0\pi^0$ is expected. 
\end{itemize}
As a by-product, we determined the cross section $\sigma_{e^+e^-\to\eta\gamma}$
at $\sqrt{s}=1$ GeV, with accuracy better than
the closer data points.
These results are encouraging also in view of the forthcoming data taking
campaign of the KLOE-2 project~\cite{KLOE2}, when both low
and high energy $e^\pm$ tagging devices will be available.

\section*{Acknowledgments}

We thank the DAFNE team for their efforts in maintaining low background
running conditions and their collaboration during all data-taking. 
We want to thank our technical staff: 
G.~F.~Fortugno and F.~Sborzacchi for their dedication in ensuring
efficient operation of the KLOE computing facilities;
M.~Anelli for his continuous attention to the gas system and detector
safety;
A.~Balla, M.~Gatta, G.~Corradi and G.~Papalino for electronics
maintenance;
M.~Santoni, G.~Paoluzzi and R.~Rosellini for general detector support;
C.~Piscitelli for his help during major maintenance periods.
This work was supported in part by EURODAPHNE, contract FMRX-CT98-0169; 
by the German Federal Ministry of Education and Research (BMBF) contract
06-KA-957; 
by the German Research Foundation (DFG), 'Emmy Noether Programme', 
contracts DE839/1-4;
by the EU Integrated Infrastructure Initiative HadronPhysics Project
under contract number RII3-CT-2004-506078;
by the European Commission under the 7th Framework Programme through
the 'Research Infrastructures' action of the 'Capacities' Programme,
Call: FP7-INFRASTRUCTURES-2008-1, Grant Agreement N. 227431;
by the Polish Ministery of Science and Higher Education through the
Grant No. 0469/B/H03/2009/37.

\section*{References}

\end{document}